\documentclass[pre,twocolumn,aps,superscriptaddress]{revtex4}

\usepackage[utf8]{inputenc}

\usepackage{array}
\usepackage{booktabs}
\usepackage{mathrsfs}
\usepackage{mathbbol}
\usepackage{multirow}
\usepackage{amsmath}
\usepackage{amsthm}
\usepackage{amssymb}
\usepackage{stmaryrd}
\usepackage{graphicx}
\usepackage{latexsym}
\usepackage{textcomp}
\usepackage{mathtools}
\usepackage{xcolor}
\usepackage[citecolor=magenta,colorlinks=true]{hyperref}
\usepackage{lineno}
\usepackage{stackrel}
\usepackage[toc,page,titletoc]{appendix}
\usepackage{bm}
\usepackage{physics}
\definecolor{mycolor1}{rgb}{0.1, 0.6, 0.6}
\usepackage{comment}
\usepackage[version=4]{mhchem}

\usepackage[nameinlink,capitalize]{cleveref}

\newcommand{\ahum}[1]{``#1''}

\newcommand{\PMD}{P_{\rm dis}}

\begin{document}

\title{Friction of a driven chain: Role of momentum conservation, Goldstone and radiation modes}

\author{Debankur Das}
\email{debankur.das@uni-goettingen.de}
\affiliation{Institut für Theoretische Physik, Georg-August-Universität Göttingen,
37073 Göttingen, Germany}

\author{Richard Vink}
\email{rvink1@gwdg.de}
\affiliation{Institut für Materialphysik, Georg-August-Universität Göttingen, 
37073 Göttingen, Germany}

\author{Matthias Kr{\"u}ger}
\email{matthias.kruger@uni-goettingen.de}
\affiliation{Institut für Theoretische Physik, Georg-August-Universität Göttingen, 
37073 Göttingen, Germany}

\date{\today}

\begin{abstract} We analytically study friction and dissipation of a driven bead in a 1D harmonic chain, and analyze the role of internal damping mechanism as well as chain length.
Specifically, we investigate Dissipative Particle Dynamics and Langevin Dynamics, as paradigmatic examples that do and do not display  translational symmetry, with distinct results: For identical parameters, the friction forces can differ by many orders of magnitude. For slow  driving, a Goldstone mode traverses the entire system, resulting in friction of the driven bead that grows arbitrarily large (Langevin) or gets arbitrarily small (Dissipative Particle Dynamics) with system size. For a long chain, the friction for DPD is shown to be bound, while it shows a singularity (i.e. can be arbitrarily large) for Langevin damping. For long underdamped chains, a radiation mode is recovered in either case, with friction independent of damping mechanism. 
For medium length chains, the chain shows the expected resonant behavior. 
At the resonance, friction is non-analytic in damping parameter $\gamma$, depending on it as $\gamma^{-1}$. Generally, no zero frequency bulk friction coefficient can be determined, as the limits of small frequency and infinite chain length do not commute, and we discuss the regimes where "simple" macroscopic friction occurs.

\end{abstract}

\maketitle

\section{Introduction}
 
Dissipation in matter is an ubiquitous phenomenon and is a key ingredient in understanding material behaviour in presence of external forces. Tailoring dissipation in solids is essential for low-friction technologies such as lubricants~\cite{xiao20172d}, 
 sonolubricity~\cite{pfahl2018universal} or wear protection~\cite{berman2013few,marian2022layered} and help in reducing energy consumption in industrial applications. Implementation of such technologies however requires understanding how microscopic mechanisms result in friction~\cite{persson1985brownian,persson1985vibrational,lee2020spatially,panizon2018analytic,sukhomlinov2021viscous,hu2020origin}. 

Recent works investigate the dependence of friction on the thickness of the involved substrate or coating \cite{marian2022layered,weber2022nanoscale,kwon2012enhanced,filleter2009friction,de2016flexible}, for which also a variety of theoretical  predictions exist~\cite{lee2023friction,benassi2010parameter}. 
Such dependence may not only yield additional ways of tuning and tailoring friction, it also yields fundamental understanding, such as where  energy is dissipated in the material, or how deep friction feels into the material. 
Several experimental studies~\cite{berman2013few,lee2010frictional, filleter2009friction, filleter2010structural, li2010substrate,kwon2012enhanced, weber2022nanoscale} have been performed to understand effects of sample thickness with contrasting results. In layered materials the friction was found to decrease with increasing number of layers~\cite{filleter2009friction,lee2010frictional,andersson2020understanding} whereas friction forces are observed to increase with increasing sample thickness in non-layered  materials~\cite{weber2022nanoscale} and  in several numerical studies~\cite{benassi2010parameter,kajita2009deep,lee2023friction}. There is thus still need for understanding.

A fundamental theoretical question concerns the appearance of friction and dissipation from integration of degrees of freedom \cite{CALDEIRA1983374,Zwanzigbook}: While the macroscopically observed laws of friction are often very simple, it is nontrivial how they emerge from microscopic dynamics~\cite{bowden2001friction,vanossi2013colloquium,bonfanti2017atomic}. In situations with finite number of particles, as is the case in numerical simulations, a microscopic damping mechanism or thermostating typically needs to be used, raising the question of the influence of such methods, and their interpretation with regards to experiments~\cite{vink2019connection,benassi2010parameter,jansen2010temperature}. 
We will focus on two paradigmatic versions of microscopic damping that find wide use. i) In Langevin damping, each atom experiences a friction force proportional to its velocity~\cite{vink2019connection, benassi2010parameter,toton2010temperature,lee2020spatially}. As these damping forces act with respect to a "solvent" medium, Langevin damping is not translationally invariant, and does not conserve momentum. ii) In  { dissipative particle dynamics} (DPD)~\cite{hoogerbrugge1992simulating, espanol1995statistical}, microscopic damping forces  are proportional to the relative velocities of  neighboring particles. This keeps translational symmetry and conserves momentum.  Although DPD  was originally designed for fluids~\cite{karniadakis2006microflows,liu2015dissipative}, it has been successfully implemented in diverse systems recently, e.g., in electron dynamics in a metal~\cite{tamm2018langevin} or biological tissues~\cite{tong2023linear}. Notably, DPD with elastic interactions corresponds to the discrete version of the Kelvin Voigt model \cite{lee2021noncontact}, which has been extensively used in theoretical analysis of nanoscale-friction on solid surfaces~\cite{persson1985brownian,persson1985vibrational,lee2021noncontact,lee2023friction}.  

In this paper, we analyze the simple textbook system of a  driven one-dimensional (1D) harmonic chain. It allows analytical treatment and yields a plethora of insights for the friction of the driven bead, which, to our knowledge, have not been reported. We study the role of microscopic damping as well as chain length, by computing the dissipated power and frequency dependent friction coefficient of the periodically driven end bead of the chain. We demonstrate that the macroscopic friction coefficient is highly sensitive to the microscopic damping mechanism. We observe the following phenomena.

i) For both Langevin and DPD cases, the limits of zero frequency and infinite chain lengths do not commute, as they correspond to fundamentally different modes. It is thus not immediately obvious, how a simple small frequency result may emerge.  

ii) For short chains, friction results from a system-spanning Goldstone mode. This mode yields a friction coefficient that grows arbitrarily large (small)  with system size for Langevin (DPD) cases. We discuss the (partly possible) mapping of these cases to the center of mass diffusion of a Rouse polymer \cite{Rouse53} (Langevin) and the shear force of a simple fluid \cite{dhont} (DPD). 

iii) For long chains, the friction for DPD is bound (Eq.~\eqref{eq:bound} below). In contrast, for Langevin, it shows a singularity on a line in parameter space where it thus gets arbitrarily large. For underdamped long chains, a radiation mode is found, with a universal friction coefficient. For DPD, this mode is always reached for small frequency, while it requires a fine tuning of parameters for Langevin damping. 

iv) Intermediate chain lengths can show resonances \cite{sarkar2019vibrational,belbasi2014anti}, which, in parameter space, are also more typically found for DPD compared to Langevin damping. At resonance, the friction coefficient of the driven bead increases with decreasing microscopic damping, so that, at this wavelength, the friction coefficient is a non-analytic function of the microscopic damping parameter.

v) A frequency (and chain length) independent friction coefficient is found only for very long chains, longer than the decay length of waves. Even under this condition, overdamped chains show non-trivial frequency dependencies, with friction coefficient depending on the square root, or inverse square root of frequency. These cases can be mapped on the monomer diffusion coefficient of a Rouse polymer and on so called vorticity diffusion of a sheared fluid \cite{dhont}, respectively.  Given that in reality, the decay length can be kilometers, the question of how a simple macroscopic friction law may arise is hardly answered in this model.

The paper is organized as follows. In Section~\ref{sec_1}, we introduce the model of a harmonic 1D chain along with its equations of motions, the boundary conditions and the damping schemes. We also present  the resulting displacement profile in presence of external driving. We introduce the macroscopic friction coefficient $\Gamma$ in Section~\ref{sec_2}. In Sec.~\ref{sec_4} and Sec.~\ref{sec_5}, we derive and analyze the resulting frequency dependent friction coefficient and analyze it in different regimes of driving frequency, damping coefficient, and chain lengths in presence of DPD damping and Langevin, respectively. Finally, we conclude the paper in Section~\ref{sec:dis} by discussing our results along with  possible experimental implications and future research directions.

\section{Model: 1D harmonic chain}
\label{sec_1}

\subsection{Harmonic chain and eigenmodes}

\begin{figure}
\centering
\includegraphics[scale=0.47]{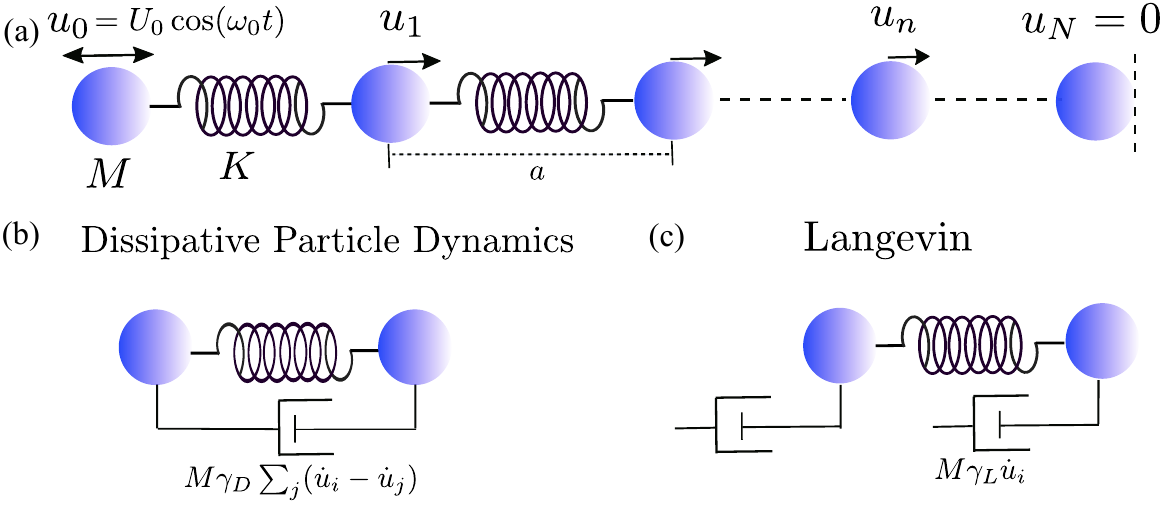}
\caption{\label{figure_model}(a) Schematic of the 1D harmonic chain composed of beads of mass $M$, connected by springs with spring constant $K$ and zero force length $a$. The $0^{\rm th}$ bead is driven externally (black arrow), i.e., it moves with a prescribed periodic protocol. Two types (b, c) of internal dissipation are considered, indicated with \ahum{dashpots}. In Langevin damping (b), dashpots connect each particle to a  background medium. In DPD (c), dashpots connect neighboring particles in the chain. 
}    
\end{figure}
We consider a 1D harmonic chain as shown in Fig.~\ref{figure_model}(a), where neighboring beads are coupled via a potential $V=\frac{K}{2}\sum_n  (u_n-u_{n-1})^2$, with a coupling strength $K$ and $u_n$  the displacement of bead~$n$ from the ground state position. With $M$ the mass of a bead, Newton's equation of motion for bead $n$ is  \cite{kardar2007statistical}
\begin{equation}
\label{eq:N2}
M \ddot{u}_n = K(u_{n-1} + u_{n+1} - 2u_n) .
\end{equation}
The displacements $\{u_n\}$
can be Fourier transformed in space and time by introducing a lattice spacing $a$, yielding $\tilde{u}_q( \omega) = \int_{-\infty}^{\infty} \sum_{n =0}^{N} e^{i(q n a - \omega t)} u_n(t) dt$. Eq.~\eqref{eq:N2} thus yields the familiar dispersion relation \cite{kardar2007statistical,kittel2005introduction} 
of the infinite length system

\begin{equation}
\omega = 2 \Omega_0 \sin(\frac{a q}{2}) \, , \quad \Omega_0 \equiv \sqrt{\frac K M} \,.
\end{equation}
We use $\Omega_0$ later as the main unit for frequency. 

\subsection{Microscopic Damping: Langevin and DPD}

Next we introduce microscopic damping and consider two paradigmatic types: Langevin and DPD damping, as shown in Fig.~\ref{figure_model}(b) and Fig.~\ref{figure_model}(c). 

\subsubsection{Langevin Damping}

In presence of Langevin damping (see Fig.~\ref{figure_model}(b)),  bead $n$ feels a damping force proportional to its velocity $\dot u_n$, quantified  by the damping coefficient $\gamma_L$, yielding 
\begin{equation}
\label{eq_Lang_motion}
M \ddot{u}_n = K(u_{n-1} + u_{n+1} - 2u_n) - M \gamma_L \dot{u}_n .
\end{equation}
Note that  $\gamma_{L}$ carries units of frequency, which allows for easy comparison to inertial contributions. The dispersion relation is obtained to be, 
\begin{eqnarray}
\label{eq:dispersionL}
4 \sin^2\left(\frac{q_La}{2}\right) = 
  \tilde{\omega}^2 + i \tilde{\omega}\tilde{\gamma}_L,
\end{eqnarray}
where we note that comparing $\omega$ and $\gamma_L$ yields over- and underdamped cases, as will be discussed below. We introduced dimensionless frequency and damping, $\tilde \omega\equiv\frac{\omega}{\Omega_0}$ and $\tilde \gamma_L\equiv\frac{\gamma_L}{\Omega_0}$. For later reference, we explicitly give the  complex wavenumber $q_L$ by solving Eq.~\eqref{eq:dispersionL} for it,
\begin{equation}
a q_L \equiv\tilde q_L= 2 \arcsin \left( \frac{1}{2}  \sqrt{\tilde \omega ^2+i \tilde \gamma_L\tilde  \omega }\right).
\label{eq_dispersion2}
\end{equation}
The displacement profile is $u_n(\omega) =  C_1 (\omega) e^{i q_L(\omega) na} +  C_2 (\omega) e^{-i q_L(\omega) na}$,
with the constants $C_1$ and $C_2$  determined from the boundary conditions introduced in subsection \ref{sec:BC} below.

The real and imaginary parts of the wave number in Eq.~\eqref{eq_dispersion2} correspond to the wavelength of modes via 
$\frac{\lambda_L}{a}\equiv \tilde \lambda_L=\frac{2\pi}{\Re \tilde q_L}$,
and the decay length 
$\frac{l_L}{a}\equiv\tilde l_L=\frac{1}{\Im \tilde q_L}$.

Notably, this model is mathematically similar to the Rouse model for a polymer chain \cite{Rouse53}. We will comment on the connections after Eq.~\eqref{eq_Gamma_lang_N} and  Eq.~\eqref{eq_Gamma_lang_omega3} below.
\subsubsection{DPD damping}

In presence of DPD damping (see Fig.~\ref{figure_model}(c)), the damping force acting on bead $n$ is proportional to its relative velocity with respect to its  neighbors. Considering only nearest neighbour damping, 
the DPD equation of motion is,
\begin{equation}
\label{eq:N2DPD}
\begin{split}
M \ddot{u}_n = \left(K+ M \gamma_D\frac{\partial}{\partial t}\right)(u_{n-1} + u_{n+1} - 2u_n).
\end{split}
\end{equation}
The damping coefficient $\gamma_D$ carries, as $\gamma_L$, units of frequency, and we use $\tilde{\gamma}_D \equiv \frac{\gamma_D}{\Omega_0}$, dimensionless. The dispersion relation for the case of DPD damping is,
\begin{equation}
\label{eq:disp_DPD}
  \sin^2\left(\frac{\tilde q_D}{2}\right) = 
  \frac{ \tilde{\omega}^2} {4(1  - i \tilde{\omega} \tilde{ \gamma}_D)} ,
\end{equation}
which shows that, comparing   $\tilde{\omega}\tilde{\gamma}_D$ to unity distinguishes over- and underdamped limits. Solving explicitly  
for $q_D(\omega)$,
\begin{equation}
aq_D\equiv\tilde q_D =  2 \arcsin \left( \frac{1}{2}  \sqrt{\frac{ \tilde{\omega}^2 }{(1  - i \tilde{\omega}  \tilde{\gamma}_D)}}\right).\label{eq_wavenumber_DPD_f}
\end{equation}
Naturally, also here, the real and imaginary parts of the wave number in Eq.~\eqref{eq_wavenumber_DPD_f} correspond to the wavelength and decay length, $\frac{\lambda_D}{a}\equiv \tilde \lambda_D=\frac{2\pi}{\Re \tilde q_D}$  and $\frac{l_D}{a}\equiv\tilde l_D=\frac{1}{\Im \tilde q_D}$.

As mentioned in the introduction, the continuum version of Eq.~\eqref{eq:N2DPD} is the Kelvin Voigt Model~\cite{lee2021noncontact}. In the absence of elastic forces, this is equivalent to the \mbox{(Navier-)Stokes} equation for a fluid \cite{dhont}. We will comment  on the connections to shearing a fluid after Eq.~\eqref{eq_Gamma_DPD_N} and Eq.~\eqref{eq_Gamma_DPD_omega_over} below.

\subsection{Boundary conditions: Strain controlled anchored chain}\label{sec:BC}
We introduce so called strain-controlled driving,  where the displacement of the bead $n=0$ is  prescribed to be
\begin{align}
u_0(t) = U_0 \cos(\omega_0 t)\label{eq:driving},
\end{align}
with $\omega_0$ the driving frequency and $U_0$ the strain amplitude. 
The other end of the chain is held fixed, i.e., {\it anchored}  so that 
\begin{eqnarray}
u_N(t) &=& 0.
\end{eqnarray}
For the monochromatic driving of Eq.~\eqref{eq:driving}, the displacement profile is monochromatic as well, 
\begin{align}
u_n(t)&=\text{Re}\{u_n(\omega_0) e^{i \omega_0 t}\},
\label{eq_disp}
\end{align}
which introduces the Fourier amplitude $u_n(\omega_0)$ of bead $n$ with $0\leq n\leq N$, $u_n(\tilde \omega_0)/U_0\equiv\tilde u_n(\tilde \omega_0)$. For the given boundary conditions, it is given by 
\begin{align}
    \tilde u_n(\tilde\omega_0)=\frac{\sin\left((N-n)\tilde q\right)}{\sin\left(N\tilde q\right)}.\label{eq:uq}
\end{align}
For the Langevin case, this is explicitly,
\begin{align}
 \tilde u_n (\tilde\omega_0)&=\frac {\sin \left(2 (N-n ) \arcsin \left(\frac{1}{2} \sqrt{ \tilde{\omega}_0  (\tilde{\omega}_0 +i \tilde{\gamma}_L )}\right)\right)}{\sin \left(2 N \arcsin \left(\frac{1}{2} \sqrt{ \tilde{\omega}_0  (\tilde{\omega}_0 +i \tilde{\gamma}_L )}\right)\right)}, 
 \label{eq_disp_Lang}
\end{align}
and for  DPD,
\begin{align}
   \tilde  u_n(\tilde\omega_0) =&  \frac{ \sin \left(2 (N-n) \arcsin \left(\frac{ \tilde{\omega}_0 }{2 \sqrt{1-i \tilde{\gamma}_D   \tilde{\omega}_0 }}\right)\right)}{\sin \left(2 N \arcsin \left(\frac{ \tilde{\omega}_0 }{2 \sqrt{1-i \tilde{\gamma}_D   \tilde{\omega}_0 }}\right)\right)}.
\label{eq_dpd_motion}
\end{align}

\section{Dissipated power and friction coefficient}
\label{sec_2}
\subsection{Dissipated Power}

The dissipated power $\PMD$ is defined as the rate of work done when driving the top bead $n=0$. It takes the general form as an average over the driving period $\frac{2\pi}{\omega_0}$,
\begin{equation}
\label{eq_diss}
\PMD = \frac{\omega_0}{2\pi}\int_0^{\frac{2\pi}{\omega_0}} dt F(t) \, \dot u_0(t)  .
\end{equation}
$\dot u_0(t)=-\omega_0U_0\sin(\omega_0 t)$ is the prescribed velocity of the end bead $n=0$ according to Eq.~\eqref{eq:driving}, and $F_0(t)$ is the force acting on that bead, i.e., the force required to exercise that motion. For the harmonic chain, the force is also monochromatic at $\omega_0$, i.e.,  $F (t) = \text{Re}\{F(\omega_0) e^{i \omega_0 t}\}$ with a complex amplitude $F(\omega_0)$. $\PMD$ may then be expressed in terms of it via
\begin{eqnarray}
\PMD = \frac{1}{2} U_0  \omega_0 \Im \{F (\omega_0)\}.
\label{eq_diss_2}
\end{eqnarray}

\subsection{Macroscopic Friction Coefficient $\Gamma$}
The quantity of interest is the friction coefficient $\Gamma(\omega_0)$ of the bead $n=0$. We will also refer to it as the macroscopic friction coefficient to distinguish from microscopic damping $\gamma$. Using the above definitions of Fourier amplitudes, it can  be extracted as the ratio 
\begin{eqnarray}
\Gamma(\omega_0) = \frac{\text{Im} \{F(\omega_0)\}}{\omega_0 U_0} ,
\label{eq_damping}
\end{eqnarray}
i.e., it is proportional to the component of $F$ in phase with the velocity of the driven bead. Comparing Eq.~\eqref{eq_damping} and Eq.~\eqref{eq_diss_2},  dissipated power and the macroscopic friction parameter $\Gamma$ have the obvious  relation, 
\begin{eqnarray}
P_{\rm dis}(\omega_0) =  \frac{1}{2}\Gamma(\omega_0) \omega_0^2 U_0^2.
\label{eq_damping_diss}
\end{eqnarray}
 We will in the following exclusively discuss the behavior of $\Gamma$, from which $\PMD$ may be extracted via Eq.~\eqref{eq_damping_diss}.
 


 \section{Friction coefficient for DPD}
 \label{sec_4}
\subsection{Force acting on driven bead}
We start by investigating the case of DPD damping. The expression for the force acting on the driven bead in presence of DPD damping is given by
\begin{align}
\frac{F}{M\Omega_0^2 U_0}\equiv \tilde F_0(\tilde\omega_0) = (1  - i  \tilde{\gamma}_D \tilde{\omega}_0 )(\tilde u_1(\tilde{\omega}_0)-\tilde u_0(\tilde{\omega}_0)).
\label{eq_dpd_force}
\end{align}
It is expressed in terms of the difference of displacements of bead 0 and bead 1, and contains a contribution from both the potential (spring) as well as the damping (dashpot) connecting these particles in Fig.~\ref{figure_model}. As the friction force, via Eq.~\eqref{eq_damping}, is given by the imaginary part of $\tilde F$, the first term in Eq.~\eqref{eq_dpd_force} will pick up the imaginary part of $\tilde u_1(\tilde{\omega}_0)$, i.e., the motion out of phase with the driven bead. The second term will pick up the real part of $\tilde u_1(\tilde{\omega}_0)-\tilde u_0(\tilde{\omega}_0)$, i.e., the part in phase with the driven bead.

We will in the following consider several  regimes of  chain length compared to wavelength and decay length.
\subsection{Short chain}  
\begin{figure*}
    \centering
    \includegraphics[scale=1.1]{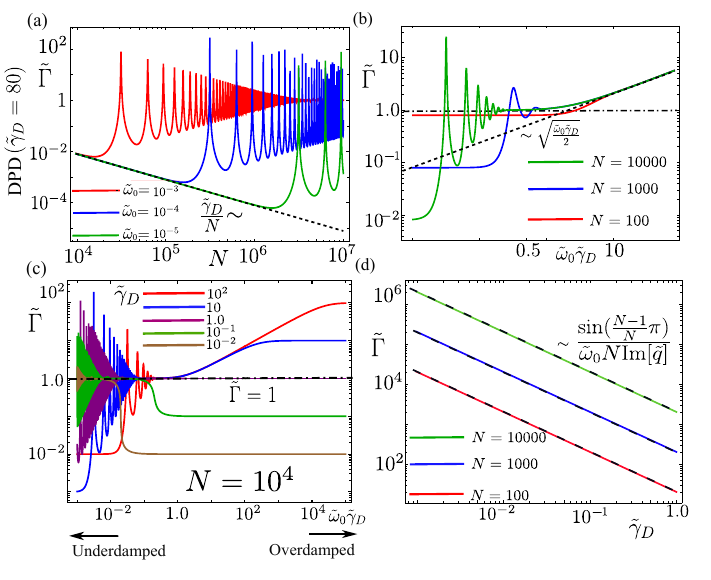}
    \caption{{\bf  DPD Damping} Macroscopic friction coefficient $\tilde{\Gamma}$ for an anchored chain when the topmost bead is driven  externally by  strain controlled driving $u_0(t) = U_0 \cos(\omega_0 t)$ in presence of DPD damping.  {\bf {\it Top}} (a)  $\tilde{\Gamma}$ vs $N$ for different driving frequencies $\tilde{\omega}_0$, (b) $\tilde{\Gamma}$ vs $\tilde{\omega}_0 \tilde{\gamma}_D$ for different chain lengths $N$, both graphs for $\tilde{\gamma}_D = 80$. Dashed lines represent the theoretically obtained forms. {\bf {\it Bottom}} (c) $\tilde{\Gamma}$ vs $\tilde{\omega}_0 \tilde{\gamma}_D$ for different values of the friction coefficient $\tilde{\gamma}_D$ for a chain of length $N=10^4$, recalling that underdamped and overdamped limits correspond to $\tilde{\omega}_0 \tilde{\gamma}_D \ll 1$ and $\tilde{\omega}_0 \tilde{\gamma}_D \gg 1$, respectively. (d) Resonant chain: Variation of $\tilde{\Gamma}$ with internal damping parameter $\tilde{\gamma}_D$  for different values of chain length $N$. We keep $N\text{Re} [\tilde{q}_D] = \pi+\mathcal{O}(\tilde\gamma_D)$  by fixing $\tilde{\omega}_0$ such that $N \arcsin[\frac{\tilde{\omega}_0}{N}] = \pi$. The dashed line represents the theoretical prediction of Eq.~\eqref{eq:res2}.
 }
    \label{fig_macro_gamma_dpd}
\end{figure*}

For a short chain, $N\ll \tilde{\lambda}_D, \tilde{l}_D$, we expand the sinus in Eq.~\eqref{eq_dpd_motion} for small arguments, to obtain the displacement in leading orders in driving frequency,    \begin{align}
    \tilde u_n(\tilde{\omega}_0)= \left(\frac{ (N-n)}{N} +\tilde{\omega}_0^2 \frac{  n(N-n) \left(2N - n \right)}{6 N} +\dots\right).
    \label{eq:un}
\end{align}
 Notably, to leading orders, $u$ is purely real: The imaginary part is of order ${\omega}_0^3$, i.e., an imaginary term of order $\omega_0$ is missing: It cancels between numerator and denominator in Eq.~\eqref{eq_disp_Lang}. The force in Eq.~\eqref{eq_dpd_force} is, up to order $\omega_0$,  thus exclusively given by the first term in Eq.~\eqref{eq:un},
\begin{eqnarray}
\tilde F_0(\tilde \omega_0) = \frac{(1 - i\tilde{\gamma}_D   \tilde{\omega}_0)}{N}.
\label{eq_dpd_force_omega_small}
\end{eqnarray}
Eq.~\eqref{eq_dpd_force_omega_small} shows two terms. The first term is real, and yields the elastic force component.  The second is imaginary, and it yields the friction coefficient $\Gamma$. Notably, both contributions vanish for large $N$ as $1/N$, which is the signature of a Goldstone mode \cite{kardar2007statistical}. In this limit, the displacement profile corresponds  to quasistatic shear \cite{dhont}, for which the relative displacement of beads shows the $1/N$ form. Notably, also the frictional component of force in Eq.~\eqref{eq_dpd_force_omega_small} vanishes as $1/N$, showing that the Goldstone mode yields vanishing friction in DPD. Explicitly, $\tilde{\Gamma} \equiv\Gamma/M \Omega_0$ follows from Eq.~\eqref{eq_damping},
\begin{eqnarray}
 \lim_{N\ll \tilde \lambda_D,\tilde l_D}\tilde \Gamma = \frac{\tilde\gamma_D }{N}.
\label{eq_Gamma_DPD_N}
\end{eqnarray}
Fig.~\ref{fig_macro_gamma_dpd} (a) shows numerical results for $\Gamma$ as a function of $N$. The figure shows the underdamped case, i.e.,   $1\gg\tilde\gamma_D\tilde\omega_0$. The used parameters also fulfil  $\tilde\gamma_D\tilde\omega_0\gg\tilde\omega_0^2$, for  which $|\tilde q_D|\ll 1$; Wave- and decay length are thus large compared to lattice spacing $a$, corresponding to the continuum limit. With $\tilde q_D\approx\tilde\omega_0$ we expect
Eq.~\eqref{eq_Gamma_DPD_N} to be obeyed for $N\alt \tilde\omega_0^{-1}$.  This is indeed confirmed by the curves in the graph, which deviate from Eq.~\eqref{eq_Gamma_DPD_N} for    $N\agt \tilde{\omega}_0^{-1}$. 

We remark that Eq.~\eqref{eq_Gamma_DPD_N} is also valid for overdamped cases, and can be mapped on the shear force felt by a simple fluid when between parallel plates of finite distance \cite{dhont}, see also the discussion section \ref{sec:dis}. 

 \subsection{Long chain}\label{sec:DPDlong}
For a long chain, we use the exponential representation of  $\sin(x)=\frac{1}{2i}(e^{ix}-e^{-ix})$. If the chain is long compared to  the decay length $l_D$, $N\gg\tilde l_D$, we can neglect the term $e^{iN a q_D}$. The dependence on $N$ cancels in Eq.~\eqref{eq_dpd_motion}, and the result becomes independent of the boundary condition at $n=N$. We obtain in this limit, 
  \begin{align}
     \tilde u_n(\tilde \omega_0) =e^{ i n \tilde q_D}= e^{-\frac{n}{\tilde l_D}}\left[i\sin\frac{2\pi n}{\tilde\lambda_D}+\cos\frac{2\pi n}{\tilde\lambda_D}\right].
     \label{eq_dpd_disp_limN}
 \end{align}
The long chain thus shows the expected behavior of an exponentially damped profile, with the meanings of $\tilde l_D$ and $\tilde\lambda_D$ becoming explicit.   This is inserted in  Eq.~\eqref{eq_dpd_force} for the force, which then leads the following expression for friction
via Eq.~\eqref{eq_damping}
\begin{align}
 \lim_{N\gg \tilde l_D}   \tilde \Gamma =  \frac{e^{-\frac{1}{\tilde l_D}}}{\tilde{\omega}_0} \sin\frac{2\pi}{\tilde\lambda_D}+\tilde{\gamma}_D\left(1-e^{-\frac{1}{\tilde l_D}}\cos\frac{2\pi}{\tilde\lambda_D}\right).
 \label{eq_Gamma_DPD_omega}
\end{align}
Eq.~\eqref{eq_Gamma_DPD_omega} gives $\tilde \Gamma$ in terms of four parameters $\tilde l_D$, $\tilde \lambda_D$,  $\tilde\omega_0$ and $\tilde\gamma_D$. These are not independent of each other, and one may express via Eq.~\eqref{eq:disp_DPD},
\begin{align}
\frac{4}{\tilde{\omega}_0^2}&=\Re\frac{1}{\sin^2(\frac{\pi}{\tilde\lambda_D}+\frac{i}{2\tilde l_D})}\label{eq:Om/om}\\
4\tilde{\gamma}_D&=-\Im\frac{1}{\sin^2(\frac{\pi}{\tilde\lambda_D}+\frac{i}{2\tilde l_D})}.
\end{align}
This shows that $\tilde\Gamma$ is determined from  $\{\tilde l_D, \tilde \lambda_D\}$, or from  $\{\tilde\omega_0,\tilde\gamma_D\}$. The mixed form of  Eq.~\eqref{eq_Gamma_DPD_omega} is advantageous because of its compactness.

Before discussing the limiting cases, we note that the second term of Eq.~\eqref{eq_Gamma_DPD_omega} is bound by $2\tilde\gamma_D$. It is also easy to show that the first term in Eq.~\eqref{eq_Gamma_DPD_omega} is bound, namely as
\begin{align}
    \frac{\Im[e^{i \tilde q_D(\tilde\omega_0,\tilde\gamma_D)}]}{\tilde\omega_0}\leq 1.
\end{align}
The friction of the long DPD chain is thus bound by
\begin{align}
 \lim_{N\gg \tilde l_D}   \tilde \Gamma\leq 1+2\tilde\gamma_D.\label{eq:bound}
\end{align}

Regarding in more detail 
we start with the underdamped limit,  $\tilde{\gamma}_D\tilde{\omega}_0\ll1$, for which generally the first term in Eq.~\eqref{eq_Gamma_DPD_omega} dominates. Taking formally $\tilde l_D\gg1$, the first term yields
\begin{eqnarray}
  \lim_{\tilde \gamma_D\tilde \omega_0\ll 1} \lim_{N\gg \tilde{l}_D} \tilde{\Gamma} = \frac{\sin{\frac{2\pi}{\tilde{\lambda}_D}}}{2\sin
 {\frac{\pi}{\tilde{\lambda}_D}}},
 \label{eq_Gamma_DPD_omega_under2}
 \end{eqnarray}
which, for allowed values of $\tilde\lambda_D\geq 2$ within the Brillouin zone varies between 0 and 1.
This result is notably independent of the damping coefficient $\tilde{\gamma}_D$, and corresponds to radiation of waves, i.e., in this case, energy is transported away, and dissipated far away from the source. Taking the continuum limit of $\tilde\lambda_D\gg1$,
\begin{eqnarray}
  \displaystyle{\lim_{\tilde\lambda_D \gg1}\lim_{\tilde\gamma_D\tilde\omega_0\ll 1} \lim_{N\gg \tilde l_D} }\tilde \Gamma = 1.
 \label{eq_Gamma_DPD_omega_under}
 \end{eqnarray}
 Notably, in this limit, $\Gamma$ is  independent of $\tilde\omega_0$, i.e., it corresponds to the limit where a simple macroscopic result for friction is found.  
 The connection to radiation is easily seen, as $\PMD=\frac{1}{2}M \Omega_0 \omega_0^2 U_0^2$, i.e., it is kinetic energy of a bead,  $\frac{1}{2} M\omega_0^2 U_0^2$, which is excited at rate $\Omega_0$, i.e., via a wave traveling with the speed of sound.
 Such radiation was also found for a small probe in a 3D system in \cite{persson1985brownian}, and its universality was discussed in \cite{lee2021noncontact}.

At the end of the Brillouin zone, i.e., as $\tilde{\lambda}_D\to 2$, this term vanishes. It corresponds to $\tilde\omega \to 2$ from below.  Of course, $\tilde\lambda_D$ cannot be smaller than 2 \cite{kardar2007statistical} (as the real part of $\arcsin$ cannot exceed $\pi/2$). For  $\tilde\omega_0>2$, thus $\tilde\lambda_D=2$, and the first term in Eq.~\eqref{eq_Gamma_DPD_omega} is identically zero, as no wave is emitted. $\tilde{\Gamma}$ is then given by the second term. As this term is small for underdamped cases, $\tilde\Gamma$ shows a sharp drop at $\tilde\omega_0=2$. For $\tilde\omega_0>2$, $\tilde{l}_D$ is finite and grows, even for underdamped cases, so that,
\begin{eqnarray}
  \displaystyle{ \lim_{\tilde\omega_0\gg 2}\lim_{\tilde{\gamma}_D\tilde{\omega}_0\ll1} \lim_{N\gg \tilde{l}_D} }\tilde{\Gamma} = \tilde{\gamma}_D,
 \label{eq_Gamma_DPD_omega_under2}
 \end{eqnarray}
This is the friction coefficient of a chain where only the driven bead moves, and the rest of the chain is at rest. This case, where no wave is emitted, is reminiscent  of evanescent waves~\cite{lee2023friction} found in 3D.  
     
In the overdamped limit, $\tilde{\gamma}_D\tilde{\omega}_0\gg1$,  the second term in Eq.~\eqref{eq_Gamma_DPD_omega} dominates.    
 In the sound wave limit, $|q_D|\ll 1$, we obtain, using $\tilde{\lambda}_D/(2\pi)=\tilde{l}_D=\sqrt{2\tilde \gamma_D/\tilde\omega_0}$,
   \begin{align}
  \displaystyle{ \lim_{|q_D| \ll 1}\lim_{\tilde{\gamma}_D\tilde{\omega}_0\gg1} \lim_{N\gg \tilde{l}_D} }\tilde{\Gamma} =   \sqrt{\frac{\tilde{\omega}_0\tilde{\gamma}_D}{2}}.
 \label{eq_Gamma_DPD_omega_over}
 \end{align}
In this limit, the friction coefficient behaves as $\sqrt{\tilde{\omega}_0}$, i.e., in the overdamped limit, there is no obvious regime of simple frequency independent macroscopic friction coefficient. 

Interestingly, $\tilde\Gamma$ in Eq.~\eqref{eq_Gamma_DPD_omega_over} equals the underdamped result multiplied by $\sqrt{\frac{\tilde{\omega}_0\tilde{\gamma}_D}{2}}$. As this term, by construction, is large in the overdamped limit, the overdamped limit generally has a large friction coefficient compared to the underdamped case.  

As a side note, we remark again that the overdamped limit can be mapped on a viscous fluid. Indeed, the result of \eqref{eq_Gamma_DPD_omega_over} corresponds to  the (shear) friction  felt by a wall moving laterally in an incompressible fluid with shear viscosity $\tilde{\gamma}_D$ \cite{dhont} (so called vorticity diffusion).

If, in the overdamped limit, we reach the end of the Brillouin zone at $\tilde\omega\approx\tilde\gamma_D$, we find
\begin{align}
  \displaystyle{\lim_{\tilde\omega_0 \gg \tilde\gamma_D}  \lim_{\tilde{\gamma}_D\tilde{\omega}_0\gg 1} \lim_{N\gg \tilde{l}_D} }\tilde{\Gamma} =   \tilde{\gamma}_D, 
 \label{eq_Gamma_DPD_omega_over2}
 \end{align}
which is again the friction of the bead when the second bead is not moving. The crossover between Eq.~\eqref{eq_Gamma_DPD_omega_over} and Eq.~\eqref{eq_Gamma_DPD_omega_over2} occurs when $\tilde\Gamma_D$ approaches $\tilde\gamma_D$ from below.
 
Fig.~\ref{fig_macro_gamma_dpd}(b) shows $\tilde\Gamma$ as a function of $\tilde\omega_0\tilde\gamma_D$ for various chain lengths $N$. With the chosen value $\tilde\gamma_D=80$, we have $\tilde\omega_0\ll1$ and $\tilde\omega_0\ll\tilde\gamma_D$ for the regimes shown in the graph, i.e., the curves correspond to the continuum limit. For the largest $N$, we both observe the radiation behavior of $\tilde\Gamma=1$ of  Eq.~\eqref{eq_Gamma_DPD_omega_under} as well as the scaling of Eq.~\eqref{eq_Gamma_DPD_omega_over}. The oscillations seen in the curves result from interference effects with reflected waves. They are not contained in the analytical formulas for a long chain, and will be discussed in the next subsection.

The different friction scaling regimes are further elucidated in Fig.~\ref{fig_macro_gamma_dpd}(c) where we have plotted $\tilde{\Gamma}$ as a function of $\tilde{\omega}_0 \tilde{\gamma}_D$ for a fixed chain length $N =1 0^4$ and different values of $\tilde{\gamma}_D$. In the underdamped limit, the oscillations are observed followed by the phonon radiation mode with $\tilde{\Gamma} = 1$ independent of the damping $\tilde{\gamma}_D$. As discussed in Eq.~\eqref{eq_Gamma_DPD_omega_under2}, $\tilde{\Gamma}$ shows a sharp drop and settles to the value $\tilde{\Gamma} = \tilde{\gamma}_D$. In the overdamped regime, $\tilde{\Gamma}$ follows $\sqrt{\tilde{\omega}_0 \tilde{\gamma}_D}/2$ in the continuum limit following Eq.~\eqref{eq_Gamma_DPD_omega_over}. Finally, the friction coefficient $\tilde{\Gamma}$ saturates to $\tilde{\gamma}_D$ (see Eq.~\eqref{eq_Gamma_DPD_omega_over2}) for large values of $\tilde{\omega}_0 \tilde{\gamma}_D $ when the decay length is $\tilde l_D \ll 1$. 

\subsection{Resonant and Anti-resonant Chain}\label{sec:res}
Eqs.~\eqref{eq:res1} and \eqref{eq:res2} below hold both for DPD and Langevin systems so that we omit indices D or L for the first part of this subsection.

The amplitude $\tilde{u}_n(\tilde{\omega}_0)$ in Eq.~\eqref{eq:uq} encodes resonances~\cite{rajasekar2016nonlinear}, occurring for  $N\Re[\tilde{q}]=p  \pi$ with integer $p$,  corresponding to   $\lambda=2\frac{N}{p}$, i.e., to positive interference of original and reflected waves. More precisely, resonances also require underdamped conditions  with $N\Im[\tilde{q}]\ll 1$. Being an underdamped phenomenon, this effect is strong for small damping, and it yields a dominant contribution from the term $\tilde F_0(\tilde\omega_0) = (\tilde u_1(\tilde\omega_0)-\tilde u_0(\tilde\omega_0))$ in both Eq.~\eqref{eq_dpd_force} as well as Eq.~\eqref{eq_lang_force}. It is the 
imaginary part of $\tilde u_1$ which contributes to friction in this case, which with the above resonance condition, and for $N\Im[\tilde{q}]\ll1$ (i.e., the wave reflects many times), is
\begin{align}
    \Im[\tilde{u}_1] =  \frac{\sin\left(\frac{N -1}{N}p\pi\right)}{N \Im[\tilde{q}]}.\label{eq:res1}
\end{align}
This yields for the friction coefficient, for both Langevin and DPD damping, in this limit,
\begin{align}
    \tilde\Gamma =  \frac{\sin\left(\frac{N -1}{N}p\pi\right)}{\tilde\omega_0 N \Im[\tilde{q}]}.
\label{eq:res2}
\end{align}
This relation shows that the friction coefficient $\tilde\Gamma$, for a given finite $N$, can grow to arbitrary large values. Notably, it diverges with $\Im[\tilde q]\to0$, providing the counter intuitive observation that smaller damping can lead larger friction. This originates from the fact that as the radiated waves are reflected a large number of times, each damping unit (bond) encounters multiple radiation waves and consequently leads to larger damping. 

On the contrary, for $(N-1)\Re[\tilde{q}]=p  \pi$ with integer $p$, the numerator of  $\tilde{u}_1(\tilde{\omega}_0)$ gets small, so that, for underdamped chains $\tilde\Gamma$ shows minima there. This situation has been termed anti-resonance \cite{belbasi2014anti,sarkar2019vibrational}.

Specifically, for DPD, regarding Eq.~\eqref{eq:disp_DPD}, the overdamped regime is found for $\tilde \omega\ll \tilde\gamma_D^{-1}$. Hence, for any value of damping $\tilde\gamma_D$, a sufficiently small value of $\tilde\omega$ renders the chain underdamped. The resonance occurs then for a sufficiently large $N$.

Fig.~\ref{fig_macro_gamma_dpd}(d) shows the variation of $\tilde{\Gamma}$ with $\tilde{\gamma}_D\ll 1$. 
For the graph, we fix $\tilde{\omega}_0$ such that $N \arcsin[\frac{\tilde{\omega}_0}{N}] = \pi$ for the different values of $N$. This corresponds to $N\text{Re} [\tilde{q}_D] = \pi+\mathcal{O}(\tilde\gamma_D)$. As expected, $\tilde{\Gamma}$ diverges as $\tilde{\gamma}_D \to 0$, following Eq.~\eqref{eq:res2}.

\section{Friction coefficient for Langevin damping}
\label{sec_5}

\subsection{General expression for force}

In presence of Langevin damping, the Fourier amplitude of the  force $F_0$ acting on the driven particle is given by 
\begin{eqnarray}
\tilde F_0(\tilde\omega_0) = (\tilde u_1(\tilde\omega_0)-\tilde u_0(\tilde\omega_0)) +i \tilde\gamma_L\tilde\omega_0  \tilde u_0(\tilde\omega_0). 
\label{eq_lang_force}
\end{eqnarray}
This force has two components, it results from the interaction with the second particle (position $u_1$) as well as the trivial part, i.e., the Langevin damping of the driven bead (via coefficient $\gamma_L$), the second term in Eq.~\eqref{eq_lang_force}. The contribution of the first term to friction is found by inserting the expressions for $u_0(\omega_0)$ and $u_1(\omega_0)$  from Eq.~\eqref{eq_disp_Lang} in Eq.~\eqref{eq_lang_force}. We again  consider several regimes of chain length compared to wavenumber in Eq.~\eqref{eq_dispersion2}.

\begin{figure*}[ht!]
    \centering
    \includegraphics[scale=1.1]{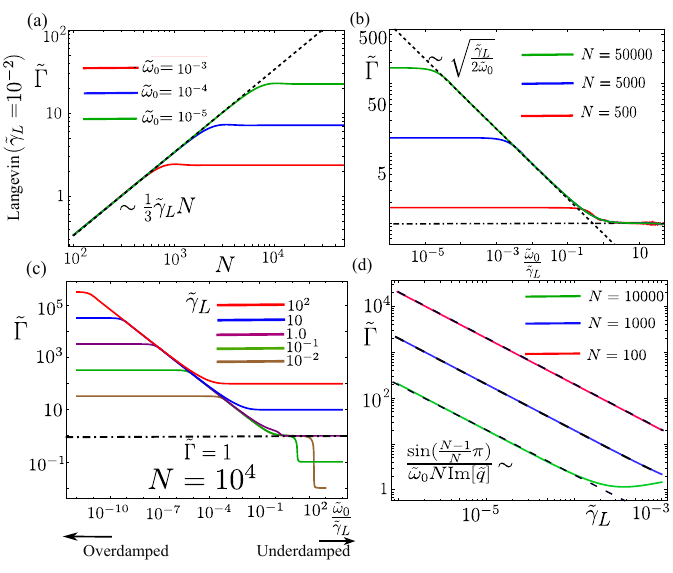}
   \caption{{\bf  Langevin Damping} Macroscopic friction coefficient $\tilde{\Gamma}$ for an anchored chain when the topmost bead is driven  externally by  strain controlled driving $u_0(t) = U_0 \cos(\omega_0 t)$ in presence of Langevin damping.  {\bf {\it Top}} (a)  $\tilde{\Gamma}$ versus $N$ for different driving frequencies $\tilde{\omega}_0$, (b) $\tilde{\Gamma}$ versus $\tilde{\omega}_0 /\tilde{\gamma}_L$ for different chain lengths $N$, both graphs for $\tilde{\gamma}_L = 10^{-2}$. The dashed lines represent the theoretically obtained forms. {\bf {\it Bottom}} (c) $\tilde{\Gamma}$ vs $\tilde{\omega}_0/ \tilde{\gamma}_L$ for different values of the friction coefficient $\tilde{\gamma}_D$ for a chain of length $N=10^4$. Recall that underdamped and overdamped limits correspond to $\tilde{\omega}_0 \gg \tilde{\gamma}_L  1$ and $\tilde{\omega}_0 \ll \tilde{\gamma}_L $, respectively. (d) Resonant chain: Variation of macroscopic $\tilde{\Gamma}$ with internal damping parameter $\tilde{\gamma}_L$  for different values of chain length $N$. The driving frequency  $\tilde{\omega}_0$ satisfies $N \arcsin[\frac{\tilde{\omega}_0}{N}] = \pi$ such that $N\text{Re} [\tilde{q}_L] = \pi+\mathcal{O}(\tilde\gamma)$ is fixed. 
   The dashed line represents the theoretical prediction in Eq.~\eqref{eq:res2}.
 }
 \label{fig_macro_gamma_lang}
\end{figure*}

\subsection{Short chain}
As for the DPD case, we start with  the limit of small $\tilde{\omega}_0$, as this corresponds to large wave- and decay lengths.  In this limit
\begin{align}
    \tilde u_n(\tilde\omega_0)= \frac{N- n}{N}+i\tilde\gamma_L\tilde\omega_0\frac{n (n^2-3 nN+2N^2)}{6 N }+\dots.
\end{align}
The real part is to leading order, as expected, identical to the DPD case. In contrast to that, here a term linear in $\tilde{\omega}_0$ exists, and it will yield the bead's friction coefficient. 
 
For the force amplitude, we obtain, 
\begin{align}
\tilde F_0(\tilde{\omega}_0) = \frac{1}{N} + 
 i  \tilde{\gamma}_L \tilde{\omega}_0\left(\frac{1}{3}  N  +\frac{ 1 }{6 N} + \frac{1}{2}  \right) +\dots. 
\end{align}
Again, the real part is the anticipated Goldstone result, vanishing as $1/N$ for large $N$. In contrast, the Goldstone mode here leads to an imaginary part which grows with $N$. Following Eq.~\eqref{eq_damping} the macroscopic friction coefficient $\Gamma$ is given by
\begin{equation}
    \lim_{N\ll \tilde{\lambda}_D, \tilde{l}_D} \tilde{\Gamma} =    \tilde{\gamma}_L   \left(\frac{1}{3}  N  +\frac{ 1 }{6 N} + \frac{1}{2}  \right).
    \label{eq_Gamma_lang_N}
\end{equation}
As already mentioned, this result if fundamentally different from the DPD result, Eq.~\eqref{eq_Gamma_DPD_N}, as it grows with $N$. This is because for Langevin damping, the excited Goldstone mode yields the dissipation of a "massive" mode as damping breaks translational symmetry: Every bead dissipates with respect to a background medium.     

Interestingly, assuming a simple (quasistatic) profile and motion of the chain, one may naively expect a prefactor of $\frac{1}{2}$  for large $N$ in Eq.~\eqref{eq_Gamma_lang_N}; the simple profile suggests that the average velocity of the beads  is half the velocity  of the driven bead. However, already to order  $\omega_0$ (which contributes to friction) $u_n$ deviates from that profile. Notably, the factor of $\frac{1}{3}$ can be understood by integrating the square of the simple profile, via $\int dx (1-x)^2=\frac{1}{3}$. Thus, the dissipated energy can be naively understood in this limit as being additive, while the friction force is not.

The numerically exact result for $\Gamma$, as a function of $N$, is plotted in Fig.~\ref{fig_macro_gamma_lang}(a) with $\tilde\gamma_L=1/100$. As $\tilde\gamma_L\gg\tilde\omega_0$ for the shown curves, the graph shows the overdamped continuum limit. In this case, $q_L\approx  \sqrt{\tilde\gamma_L \tilde\omega_0}  (1+i)/\sqrt{2}$. We expect Eq.~\eqref{eq_Gamma_lang_N} to hold for $N\ll 1/|\tilde q_L|$, and indeed, the curves in the graph deviate at around   $N \approx \sqrt{1/\tilde\omega_0\tilde\gamma_L}$.

Finally, we remark that the friction of  Eq.~\eqref{eq_Gamma_lang_N} is related to the center of mass diffusion of a Rouse polymer \cite{Rouse53}. The additional factor of $1/3$ is due to the anchoring, and is absent for a chain with a free end.

\subsection{Long chain}
If the chain is long, $N\gg\tilde l_L$, we repeat the steps of subsection \ref{sec:DPDlong}, and expand the sinus for large imaginary argument, 
\begin{align}
\tilde u_n(\tilde \omega_0) =e^{ i n \tilde{q}_L}= e^{-\frac{n}{\tilde{l}_L}}\left[i\sin\frac{2\pi n}{\tilde{\lambda}_L}+\cos\frac{2\pi n}{\tilde{\lambda}_L}\right].
     \end{align}
This is inserted in  Eq.~\eqref{eq_lang_force} for the force. Here, only the imaginary part of $u_1$ contributes, so that, 
via Eq.~\eqref{eq_damping}
\begin{align}
   \lim_{N\gg \tilde{l}_L} \tilde{\Gamma} =   \tilde{\gamma}_L+  \frac{e^{-\frac{1}{\tilde{l}_L}}}{\tilde{\omega}_0} \sin\frac{2\pi}{\tilde{\lambda}_L}.
 \label{eq_Gamma_lang_omega}
\end{align}
Let us be reminded that the first term in Eq.~\eqref{eq_Gamma_lang_omega} is the trivial contribution of the damping of the driven bead. We may express the amplitude of the second term via Eq.~\eqref{eq:dispersionL} as
\begin{align}
\label{eq:LangOm/om}
\frac{4}{\tilde{\omega}_0^2}&=\frac{1}{\Re\left[\sin^2(\frac{\pi}{\tilde{\lambda}_L}+\frac{i}{2\tilde{l}_L})\right]},
\end{align}
showing that $\tilde\Gamma$ is fully determined  from either $\tilde l_D$ and $\tilde \lambda_D$, or from  $\tilde\omega_0$ and $\tilde\gamma_D$.

Before discussing specific limits, let us ask whether $\tilde\Gamma$ is bound as for the DPD case in Eq.~\eqref{eq:bound}. With Eq.~\eqref{eq:LangOm/om}, we have  $\text{Re}[\sin\tilde{q}_L/2] \geq \Im[\sin\tilde{q}_L/2] \geq 0$ and   $\tilde\omega_0^2=4(\Re[\sin(\tilde q_L/2)]^2- \Im[\sin(\tilde q_L/2)]^2)$, and we obtain,
\begin{align}
\Gamma = \gamma_L +\frac{\Im[e^{i\tilde q_L}]}{2\sqrt{\Re[\sin(\tilde q_L/2)]^2 - \Im[\sin(\tilde q_L/2)]^2}} \label{eq:nbound}
\end{align}
We see that the denominator can get arbitrarily close to zero for $\Re[\sin(\tilde q_L/2)]^2 \to \Im[\sin(\tilde q_L/2)]^2$. Because the numerator stays finite in this limit, it corresponds to a pole, i.e., $\tilde\Gamma$  can get arbitrarily large. This is a pronounced difference to the DPD case, where $\tilde\Gamma$ for the long chain is bound according to Eq.~\eqref{eq:bound}. There is however a (trivial) bound also here, but it is a bound from below \begin{align}
    \lim_{N\gg \tilde l_D}\tilde\Gamma\geq \gamma_L.
    \label{eq:boundL}
\end{align}

As for DPD, we start with the underdamped case, which is here given for $\tilde{\omega}_0\gg \tilde{\gamma}_L$. Using formally $\tilde l_L \gg 1$, we have 
\begin{eqnarray}
   \displaystyle{\lim_{\tilde{\omega}_0\gg\tilde{\gamma}_L} \lim_{N\gg \tilde{l}_D} }\tilde{\Gamma} = \tilde{\gamma}_L+ \frac{\sin{\frac{2\pi}{\tilde{\lambda}_L}}}{2\sin{\frac{\pi}{\tilde{\lambda}_L}}},
 \label{eq_Gamma_lang_omega_under2}
 \end{eqnarray}
 which also agrees with the DPD case, Eq.~\eqref{eq_Gamma_DPD_omega_under2}, up to the first term. As for DPD, in the sound wave limit, i.e., $|\tilde \lambda_L|\gg 1$,   and 
\begin{align}
   \lim_{|\tilde q_L|\ll 1}\lim_{\tilde{\omega}_0\gg\tilde{\gamma}_L}\lim_{N\gg \tilde{l}_L} \tilde{\Gamma} = \tilde\gamma_L +1.
 \label{eq_Gamma_lang_omega2}
\end{align}
This result, corresponding to the radiation mode, is also identical to the result from DPD damping in the underdamped limit (up to the trivial term). As the second term in Eq.~\eqref{eq_Gamma_lang_omega2} is independent of friction coefficient, it should indeed be reached by either case. Notably, only for $\tilde\gamma_L\ll1$, the radiation term dominates friction, which is a condition that is additional to the one of being underdamped.

 The end of the Brillouin zone, i.e., $\tilde \lambda_L=2$ is reached for $\tilde\omega_0\approx 2$. At this point, the second term in Eq.~\eqref{eq_Gamma_lang_omega_under2} vanishes, and $\tilde{\Gamma}=\tilde{\gamma}_L$ for $\tilde\omega_0\agt 2$.

The overdamped limit corresponds to $\tilde{\omega}_0\ll\tilde{\gamma}_L$. In the continuum limit, $\tilde{\lambda}_L/(2\pi)=\tilde{l}_L=  \sqrt{2/\tilde{\gamma}_L \tilde{\omega}_0}$, and we have, 
\begin{align}
  \lim_{1\ll\tilde{\lambda}_L,\tilde{l}_L}\lim_{\tilde{\omega}_0\ll\tilde{\gamma}_L} \lim_{N\gg \tilde{l}_L} \tilde{\Gamma} =  \sqrt{\frac{\tilde{\gamma}_L}{2\tilde{\omega}_0}}=\frac{1}{2}\tilde\gamma_L \tilde l_L. 
 \label{eq_Gamma_lang_omega3}
\end{align}
 $\tilde{\Gamma}$ in Eq.~\eqref{eq_Gamma_lang_omega3} shows non-trivial behavior, as it grows as an inverse square root of $\tilde{\omega}_0$. As anticipated around Eq.~\eqref{eq:nbound}, $\tilde\Gamma$ can grow without bounds, and we see  now  how the mentioned pole is approached. As long as $\tilde\l_D\ll N$, $\tilde\Gamma$ grows with $\tilde l_L$, as the number of particles contributing to dissipation grows (second equality in Eq.~\eqref{eq_Gamma_lang_omega3}). As $\tilde l_L \sim 1/\sqrt{\tilde \omega_0}$ in the considered case, the friction thus diverges as frequency goes to to zero in the seen fashion.  

We remark that the result of Eq.~\eqref{eq_Gamma_lang_omega3} can be translated to the short time diffusion of a segment in a Rouse  polymer \cite{Rouse53}. In time domain, $\tilde\Gamma$ of Eq.~\eqref{eq_Gamma_lang_omega3} yields  $\tilde{\Gamma}(t) \sim t^{3/2}$ and the mean square displacement (MSD) $\sim t^{1/2}$. We omit here a detailed translation from friction to diffusion kernel \cite{Caspers23}. 

In the overdamped limit, the end of Billouin zone is reached for $\tilde{\omega}_0\tilde\gamma_L\approx 8$ , and the  second term in Eq.~\eqref{eq_Gamma_lang_omega} vanishes, and
\begin{align}
  \lim_{\tilde{\omega}_0\ll\tilde{\gamma}_L} \lim_{N\gg \tilde{l}_L} \tilde{\Gamma} = \tilde{\gamma}_L, 
 \label{eq_Gamma_lang_omega2}
\end{align}
which, as for the DPD chain, is dominated by the bare friction of the driven bead.

Figure \ref{fig_macro_gamma_lang}(b) shows numerically exact results for $\tilde{\Gamma}$ as a function of $\tilde{\omega}_0/\tilde{\gamma}_L$, for different values of $N$. Notably, for a given $N$, Eq.~\eqref{eq_Gamma_lang_omega} holds for $\tilde\omega_0\gg \sqrt{2}/N$. It deviates for smaller frequencies, as the decay length $\tilde l_L$ then  exceeds $N$.

The different regimes are summarized in   Fig.~\ref{fig_macro_gamma_lang}(c) which shows $\tilde{\Gamma}$ as a function of $\tilde{\omega}_0 /\tilde{\gamma}_L$ for a fixed chain  length $N =1 0^4$ and different values of damping parameter $\tilde{\gamma}_L$. When $\tilde{\omega}_0 \ll \tilde{\gamma}_L$ (the overdamped limit), $\tilde{\Gamma}$ exhibits a regime following  Eq.~\eqref{eq_Gamma_lang_omega} followed by saturation as one decreases $\tilde{\omega}_0$ beyond  $l_L \approx N$. The radiation mode ($\tilde{\Gamma} =1$) is observed at the onset of the underdamped limit for $\tilde\gamma_L\ll 1$ followed by the regime corresponding to the end of the Brillouin zone, where the friction coefficient is dominated by the bare friction of the driven bead, $\tilde{\Gamma} = \tilde{\gamma}_L$.

\subsection{Resonant chain}
The discussion of resonant damping in Sec.~\ref{sec:res} is analogous  here, i.e., Eqs.~\eqref{eq:res1} and \eqref{eq:res2} apply. Resonances are not observed in Fig.~\ref{fig_macro_gamma_lang} due to choice of parameters. In  Fig.~\ref{fig_macro_gamma_lang}, the regime of intermediate chain lengths coincides with the overdamped regime. 

 The resonant behaviour requires $\tilde{\gamma}_L \ll \tilde{\omega}_0\approx\pi/N$, i.e., a sufficiently small value of $\tilde\gamma_L$. Fig.~\ref{fig_macro_gamma_lang}(d) shows $\tilde{\Gamma}$ as a function of $\tilde{\gamma}_L$ in the resonant chain regime. Similar to the DPD case,  $N\text{Re} [\tilde{q}_L] = \pi+\mathcal{O}(\tilde\gamma_L)$ is kept fixed by choosing $\tilde{\omega}_0$ such that $N \arcsin[\frac{\tilde{\omega}_0}{N}] = \pi$. $\tilde{\Gamma}$ is observed to diverge as one approaches $\tilde{\gamma}_L \to 0$ in the resonant regime.

Interestingly, in contrast, for DPD, occurrence of \mbox{(anti-)resonances} requires $\tilde{\gamma}_D\tilde\omega \ll 1$ and  $\tilde{\omega}_0\approx\pi/N$, i.e., they can be observed for any value of damping parameter $\gamma_D$, for sufficiently large chain length $N$ and small driving frequency $\tilde{\omega}_0$.

 \section{Discussion and Outlook}
\label{sec:dis}
We have presented exact results for  the dissipation of a simple 1D harmonic chain under oscillatory driving, for two distinct types of dissipation mechanisms: Langevin damping and  Dissipative Particle Dynamics (DPD) damping. What have we learned?

As a main insight, the results from DPD and Langevin damping drastically differ. They  differ quantitatively, by orders of magnitude, and qualitatively, via different scaling behaviors. This insight manifests the importance of momentum conservation for non-equilibrium phenomena like friction. This insight is also important for modeling friction analytically and in computer simulations, in the desire to compare to experiments.  

For DPD damping, a short chain shows a small friction coefficient, $\tilde \Gamma = \frac{\tilde\gamma_D }{N}$, since relative  atomic motion is small. A chain that is long compared to absorption length obeys a bound, $\tilde\Gamma\leq 1+2\tilde\gamma_D$. The long chain further  shows the universal result of $\tilde\Gamma=1$ in the underdamped continuum limit, which is reached for small frequencies. For increasing driving frequency $\tilde\omega_0$, this behavior is cut off by the end of the Brillouin zone, where  $\tilde\Gamma=\tilde\gamma_D$ is approached. For $\tilde\gamma_D>1$, an intermittent overdamped regime exists with  $\tilde\gamma_D\sim\sqrt{\omega_0}$. Intermediate size chains show resonances where $\tilde\Gamma\sim \tilde\gamma_D^{-1}$  grows arbitrarily large. 

For Langevin damping, the short chain shows $\tilde \Gamma = \frac{N\tilde\gamma_L }{3}$, i.e., the Goldstone mode yields a friction growing with system size. A long, underdamped, continuum chain shows similarities to the DPD case. As it is bound from below, $\tilde\Gamma\geq \tilde\gamma_L$,  the universal value of unity from phonon radiation can only be seen for $\tilde\gamma_L\ll1$. For small $\omega_0$, the long chain becomes overdamped, and shows  $\tilde\gamma_D\sim1/\sqrt{\omega_0}$, in contrast to the universal small frequency behavior of the long DPD chain.  Resonances for intermediate chain lengths only occur for very small values of $\tilde\gamma_L$.   

Where are regimes of simple, frequency-independent friction? For both damping cases, these occur for short chains, where, however, friction depends on system size. For long chains, frequency- and system size independent friction plateaus are found. In realistic situations one may expect a spectrum of frequencies to be excited, which for the mentioned plateaus yields a simple superposition.         
For DPD, long chains generally approach the simple universal friction for small frequency. For the Langevin case, the small frequency limit does not show simple friction, which is thus harder to find in that case. 

In realistic solids, wavelength and decay length typically exceed meters for frequencies in the range of a few kilohertz (see Ref.~\cite{lee2023friction} for details). It may thus be hard to find the long chain limit for such materials in experiments. It is in this regard interesting to consider the role of geometry.  In contrast to the 1D system studied here, Refs.~\cite{persson1985brownian, lee2023friction} investigate a 3D system perturbed by a small (point) source for which evanescent waves exist. When evanescent waves dominate friction, bulk behavior is approached much faster, i.e., for sample heights in the range of nanometers \cite{lee2023friction}. This hints to the importance of geometry and dimensionality. 

Future work can investigate other geometries  ~\cite{liefferink2021geometric} and nonlinear interactions~\cite{braun1996mobility}.  Another avenue that can be explored is the dissipation behaviour in presence of more complicated damping  mechanisms~\cite{lucke1956ultrasonic,kishore1968acoustic}.

\section{Conflict of Interest}

The authors declare no conflict of interest.

\section{Data Availability Statement}

The datasets generated and analysed during the current study are available from the authors on request.

\section{acknowledgments}

We thank Cynthia A. Volkert and Kim Lambert  for fruitful discussions and Cynthia A. Volkert for a critical reading of the manuscript. This work was supported by the Deutsche Forschungsgemeinschaft (DFG, German Research Foundation) -- 217133147/SFB 1073, project A01.
\paragraph*{Author Contributions:} D.D. developed the theory and analyzed the different scaling regimes; R.V. initiated the study and derived the scaling laws in the quasi-static limit; M.K.  supervised the research and analyzed the scaling regimes with D.D.;  M.K. and D.D. wrote the paper.

\bibliographystyle{unsrt}
\bibliography{friction1d.bib}

\end{document}